\documentclass[aip,apl,reprint]{revtex4-1}

\usepackage{graphicx}
\usepackage{dcolumn}
\usepackage{bm}
\usepackage{hyperref}
\usepackage{longtable}
\usepackage{siunitx}
\usepackage{multirow}
\usepackage{booktabs}
\usepackage{xcolor}
\usepackage{ulem}

\begin{document}

\title{Unidirectional Distributed Acoustic Reflection Transducers for Quantum Applications}

\author{\'E. Dumur}
\affiliation{Institute for Molecular Engineering, University of Chicago, Chicago IL 60637, USA}
\affiliation{Institute for Molecular Engineering and Materials Science Division, Argonne National Laboratory, Argonne, IL 60439, USA}
\author{K. J. Satzinger}
\altaffiliation[Present address: ]{Google, Santa Barbara, California 93117, USA}
\affiliation{Department of Physics, University of California, Santa Barbara CA 93106, USA}
\affiliation{Institute for Molecular Engineering, University of Chicago, Chicago IL 60637, USA}
\author{G. A. Peairs}
\affiliation{Department of Physics, University of California, Santa Barbara CA 93106, USA}
\affiliation{Institute for Molecular Engineering, University of Chicago, Chicago IL 60637, USA}
\author{Ming-Han Chou}
\affiliation{Institute for Molecular Engineering, University of Chicago, Chicago IL 60637, USA}
\affiliation{Department of Physics, University of Chicago, Chicago IL 60637, USA}
\author{A. Bienfait}
\affiliation{Institute for Molecular Engineering, University of Chicago, Chicago IL 60637, USA}
\author{H.-S. Chang}
\affiliation{Institute for Molecular Engineering, University of Chicago, Chicago IL 60637, USA}
\author{C. R. Conner}
\affiliation{Institute for Molecular Engineering, University of Chicago, Chicago IL 60637, USA}
\author{J. Grebel}
\affiliation{Institute for Molecular Engineering, University of Chicago, Chicago IL 60637, USA}
\author{R. G. Povey}
\affiliation{Institute for Molecular Engineering, University of Chicago, Chicago IL 60637, USA}
\affiliation{Department of Physics, University of Chicago, Chicago IL 60637, USA}
\author{Y. P. Zhong}
\affiliation{Institute for Molecular Engineering, University of Chicago, Chicago IL 60637, USA}
\author{A. N. Cleland}
\affiliation{Institute for Molecular Engineering, University of Chicago, Chicago IL 60637, USA}
\affiliation{Institute for Molecular Engineering and Materials Science Division, Argonne National Laboratory, Argonne, IL 60439, USA}

\begin{abstract}
Recent significant advances in coupling superconducting qubits to acoustic wave resonators has led to demonstrations of quantum control of surface and bulk acoustic resonant modes as well Wigner tomography of quantum states in these modes \cite{OConnell2010, Chu2017, Satzinger2018}. These advances were achieved through the efficient coupling afforded by piezoelectric materials combined with GHz-frequency acoustic Fabry-Perot cavities.
Quantum control of \textit{itinerant} surface acoustic waves appears in reach, but is challenging due to the limitations of conventional transducers in the appropriate GHz-frequency band. 
In particular, GHz-frequency unidirectional transducers would provide an important addition to the desired quantum toolbox, promising unit efficiency with directional control over the surface acoustic wave emission pattern. Here we report the design, fabrication and experimental characterization of unidirectional distributed acoustic reflection transducers (DARTs) demonstrating a high transduction frequency of \SI{4.8}{\giga\hertz} with a peak directivity larger than \SI{25}{\decibel} and a directivity greater than \SI{15}{\decibel} over a bandwidth of \SI{17}{\mega\hertz}. A numerical model reproduces the main features of the transducer response quite well, with ten adjustable parameters (most of which are constrained by geometric and physical considerations). This represents a significant step towards quantum control of itinerant quantum acoustic waves.
\end{abstract}

\maketitle

Surface acoustic waves are typically launched and received using interdigital transducers (IDTs), comprising parallel metal fingers evenly spaced on a piezoelectric substrate. The center frequency $f_0$ of the transducer corresponds to an acoustic wavelength $\lambda_0$ twice the lateral spacing $p$ of the fingers, where the frequency and wavelength are related by $f_0 = v_e/\lambda_0$, where $v_e$ is the  effective surface velocity in the transducer structure. The transducer bandwidth scales inversely with the number $N$ of transducer fingers \cite{Morgan2007}. These structures have been used for many years, providing a convenient, inexpensive, and highly flexible approach to integrating acoustic-wave resonators, delay lines and pulse shaping with semiconductor electronics. Recently, in a pioneering experiment\cite{Gustafsson2014}, a superconducting qubit was coupled to surface acoustic waves (SAWs) using an IDT transducer, the circuit fabricated on a GaAs substrate with a center frequency of about 5 GHz. This experiment allowed the low-temperature observation of quantum effects mediated by SAW phonons, the quanta of mechanical vibrations, coupled to the qubit through the IDT structure. More recently, other advances in the quantum control of surface acoustics have been demonstrated\cite{Manenti2017, Moores2017, Satzinger2018}, showing promise for the expansion of this area of quantum physics.

A general challenge associated with uniform IDTs is that a voltage tone applied to the IDT will result in symmetric, oppositely-directed SAWs being emitted from the two ends of the IDT, effectively acting as a three-port device, with one electrical and two acoustic channels. This is appropriate for resonant structures where the IDT is placed between a pair of Fabry-Perot-like acoustic mirrors \cite{Morgan2007, Manenti2017, Satzinger2018, Moores2017}, but leading to the loss of half the acoustic power when used for example in a delay-line geometry. For this reason, unidirectional transducers (UDTs) are typically used to preserve acoustic power, using a more complex finger geometry or multiple phase-controlled electrical signals to emit and receive waves in a single direction. In the limit of very high directivity, a UDT can be viewed as a two-port device with one electrical and one acoustic channel, transducing signals between the two modes without loss. Future quantum applications would greatly benefit from high-directivity UDTs working in the GHz frequency band.

SAW directionality in UDTs is usually achieved using variants of IDT finger geometries designed to give constructive interference for waves emitted in one direction, and destructive interference for those in the opposite direction. Multi-phase UDTs offer very high directivity\cite{Morgan2007} but have complex finger geometries and require two or more phased voltage sources, making both fabrication and operation challenging. By contrast, single-phase UDTs (SPUDTs) achieve reasonable directivity using only a single voltage source, by using an asymmetric finger geometry. Since their introduction in 1976 \cite{Hanma1976}, these transducers have been implemented in a number of different formats \cite{Hartmann1982, Lewis1983, Kodama1986, Hode1990, Yamanouchi1992, Garber1994, Lehtonen2003, Martin2011}.

One interesting approach to SPUDTs, termed a distributed acoustic reflection transducer (DART) \cite{Kodama1986, Morgan2007}, is a design that achieves high directivity by shifting the center of transduction from the center of reflection by a distance $3 \lambda_0/8$, yielding the desired oppositely-directed constructive and destructive interference. Traditionally these have only been used at lower frequencies, below \SI{100}{\mega\hertz}, with isolated examples up to \SI{500}{\mega\hertz}\cite{Garber1994, Martin2011}. The great simplicity of fabrication and operation makes these devices appealing for quantum operations if they can be made to work at higher (GHz) frequencies, compatible with superconducting qubits. Here we demonstrate a DART design that works well at frequencies approaching \SI{5}{\giga\hertz}, an order of magnitude increase from prior work.

\begin{figure}[ht]
\centering
\includegraphics[width=0.99\linewidth]{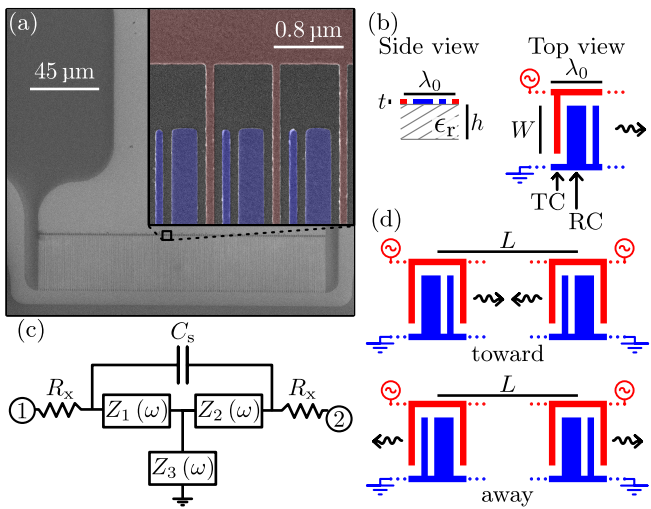}
\caption{(a) Scanning electron micrographs of a DART, with inset showing the drive and ground electrodes, driven electrodes in red and grounded electrodes in blue.
(b) Side and top view of a unitary DART cell (thickness $t$ exaggerated).
(c) Complete equivalent electrical circuit of a delay line with $Z_{1,2,3} \left(\omega\right)$ the frequency-dependent impedances as predicted by the RAM and COM theories (see main text), with embedding resistance $R_\mathrm{x}$ and capacitance $C_\mathrm{s}$.
(d) Schematic ``toward'' and ``away'' DART configurations.}
\label{fig-1}
\end{figure}

We designed and measured various DART designs, measuring single transducers as well as paired transducers in a delay line configuration, to fully characterize the transducer properties. The standard design is shown in Fig.~\ref{fig-1}, comprising driven transducer fingers of width $\lambda_0/8$ spaced by $\lambda_0$, interspersed with a $3 \lambda_0/8$- and a $\lambda_0/8$-wide pair of grounded fingers with $\lambda_0/8$ inter-finger spacing. We pattern the UDTs as a single layer of aluminum of thickness $t \approx \SI{28}{\nano\meter}$ on top of a single-side polished LiNbO\textsubscript{3} wafer of thickness $h = \SI{500}{\micro\meter}$ and room-temperature permittivity $\epsilon_\mathrm{r} = 56$. The aluminum film is lift-off patterned by electron-beam lithography, using a polymethylmethacrylate (PMMA) bilayer composed of a \SI{100}{\nano \meter} thick 495 kD weight bottom layer and a \SI{100}{\nano \meter}  thick 950 kD weight top layer. A \SI{10}{\nano\meter} thick layer of Au is thermally evaporated on the top PMMA layer to reduce charging effects. The DART geometry is kept constant along the entire transducer length, with center wavelength $\lambda_0 = \SI{800}{\nano\meter}$, corresponding to a center frequency $f_0$ just below \SI{5}{\giga\hertz}.

We designed arrays of transducers in delay-line configurations, each delay line containing a pair of DARTs, typically with a center-to-center distance $L = \SI{400}{\micro\meter} = 500\,\lambda_0$. In a typical test set of devices, we would pattern an array of samples with different transducers apertures $W = \{18, 21, 24\}~\SI{}{\micro\meter}$ and different numbers of repeat cells, $N = \{125, 130, 135\}$. Each set of transducers was fabricated in two delay-line configurations, one where the DARTs are designed to emit towards one another, and one in the opposite configuration, where they emit away from one another (Fig.~\ref{fig-1}(d)).

\begin{figure*}
\includegraphics[]{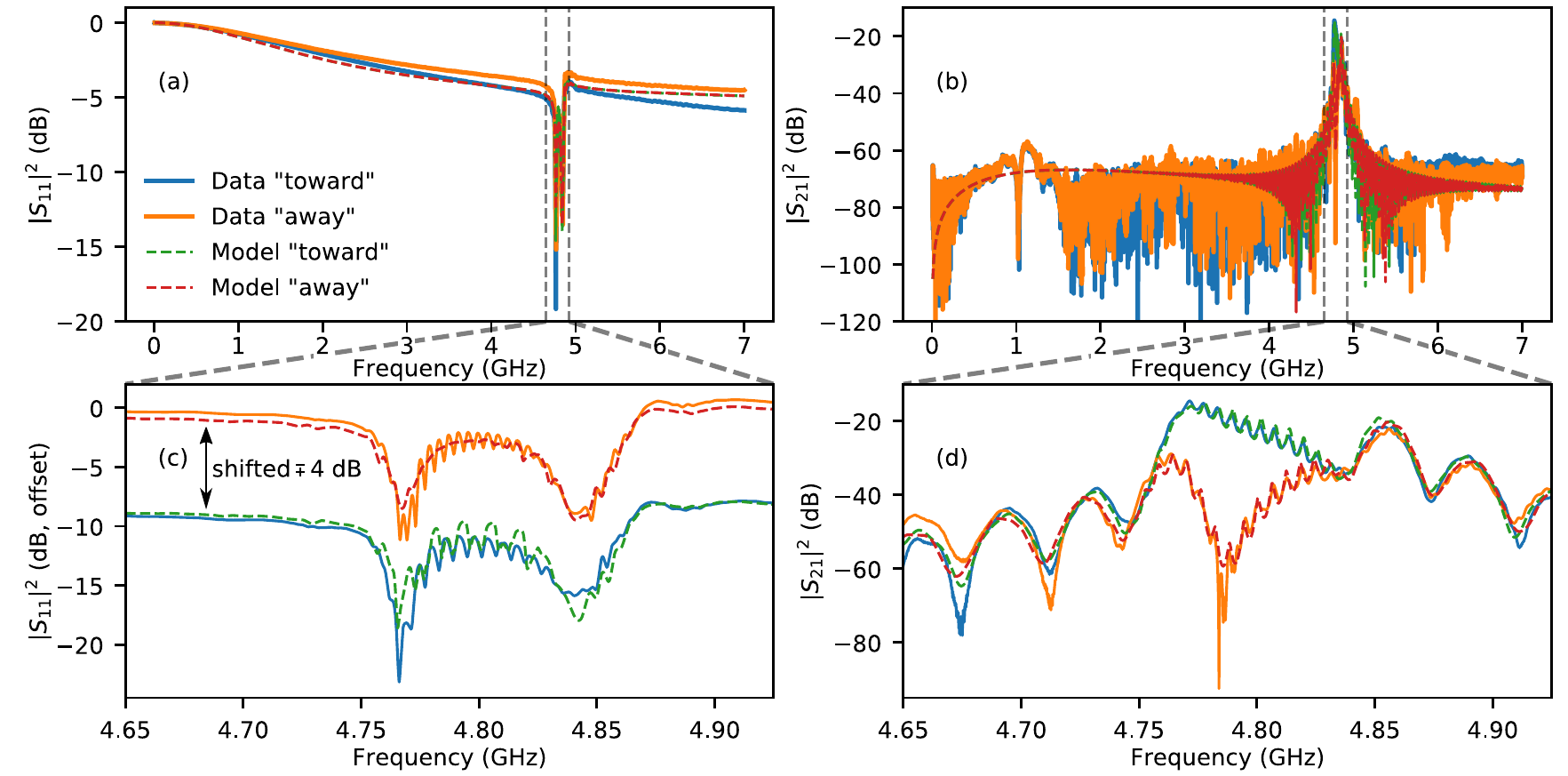}
\caption{DART paired response ($S_{11}$ and $S_{21}$) as a function of frequency, for the "towards" and "away" configurations, using solid lines, compared with numerical modeling using dashed lines. The two configurations have similar response over a broad frequency range (a-b) except near the center design frequency (c-d), where the DART directivity is manifested.}
\label{fig-2}
\end{figure*}

Measurements were made at room temperature with a vector network analyzer, using a calibrated microwave probe station. We measured both reflection $S_{11} \cong S_{22}$ and transmission $S_{21} \cong S_{12}$ between pairs of DARTs in a delay-line configuration, with typical room-temperature results shown in Fig.~\ref{fig-2}.

For the transmission measurements, we show both the ``towards'' and ``away'' configurations, allowing us to extract the transducer directivity, a measure of the directionality of the acoustic power emission.

Over a broad frequency range, the reflection $S_{11}$ and transmission $S_{21}$ show the expected behavior, which away from the design frequency $f_0$ is dominated in reflection by the interdigital capacitance of the transducers, and in transmission by the stray electrical coupling between the two transducers. For our design, the stray electrical coupling is less than \SI{-60}{\decibel} for most of the frequency range of the measurement from \SI{10}{\mega\hertz} to \SI{7}{\giga\hertz}. Near the design center frequency $f_0 = \SI{4.8}{\giga\hertz}$, we see the expected detailed frequency response for both reflection and transmission. There is a pronounced difference in the transmission for the ``towards'' and ``away'' transducer configurations, as expected for these directional transducers.

The detailed response near resonance is accurately captured by a model that takes into account the properties of the LiNbO$_3$ wafer, the aluminum electrodes and the microwave measurement circuitry. A schematic of the equivalent circuit model is shown in Fig.~\ref{fig-1}.~(c), including the transducers and the delay line equivalent impedances $Z_{1-3}$, and an approximate embedding circuit comprising a series resistance $R_x \approx \SI{15}{\ohm}$ associated with the electrodes, and a stray capacitive coupling $C_s \approx \SI{0.9}{\pico\farad}$ between the two DARTs. These two parameters provide the broad frequency response for both $S_{11}$ and $S_{21}$, away from the design frequency $f_0$.

To model the details of the DART response, we use the quasi-static approximation \cite{Morgan2007} to calculate the surface wave power as well as the electrical charge density and the DART capacitance. The wide $3\lambda_0/8$ grounded electrode of the DART is modeled as two electrodes, spaced by $\lambda_0/8$, an approximation that has been shown to be valid at lower frequencies \cite{Dufilie1995}. The effective wave velocity $v_\mathrm{e}$ is estimated using a first-order expansion developed for single finger pair IDTs and used here as an approximation\cite{Morgan2007}, where $v_\mathrm{e} = v_\mathrm{f} + \Delta v_\mathrm{e} + \Delta v_\mathrm{m}$, with $v_\mathrm{f}$ being the free-surface SAW velocity, $\Delta v_\mathrm{e}$ the velocity change due to electrical loading, and $\Delta v_\mathrm{m}$ the velocity change due to mechanical loading. The reflection, transduction, conductance and susceptance of the DARTs are modeled through the coupling-of-modes (COM) theory \cite{Morgan2007}, which models the DART cells as equally-spaced point contacts that couple the SAW modes traveling in one direction to the modes traveling in the opposite direction, through the electrical current flowing in the microwave circuit. The COM theory has two critical parameters, the transduction and reflection of each DART cell. Following Ref.~\onlinecite{Morgan2007}, we use the reflective array method (RAM) to relate the transduction to the surface wave power and electrostatic charge density; while the RAM theory was not developed for DART-style transducers, elsewhere it has been shown that this is a reliable modeling approach \cite{Morgan1998}. The reflection is kept as a fitting parameter, which we restrict to be imaginary-valued, equivalent to assuming that transmission loss in the substrate dominates over transducer-associated losses.

In the reflection signal near resonance, the response consists of two local minima near \SI{4.77}{\giga\hertz} and \SI{4.84}{\giga\hertz}, separated by a broader local maximum.
This is a characteristic of UDT-style transducers\cite{Morgan2007}.
The reflection minima are at the frequencies where transmission is maximal, and the small ${\sim} \SI{6}{\mega\hertz}$ ripples  between the minima are due to interference between single and triple transit signals between the two transducers.

In the transmission signal near resonance, the overall signal level increases by more than \SI{40}{\decibel} over the background stray coupling, with fine ripples due to interference effects within each transducer. The most striking feature is the strong difference in transmission between the ``toward'' and ``away'' DART configurations, due to the strong directivity of the transducer design. This directivity-dominated feature is more than \SI{30}{\decibel} over a span of about \SI{15}{\mega\hertz} centered at \SI{4.785}{\giga\hertz}. The asymmetry of the response about the operating frequency is thought to be due to SAW velocity dispersion and the frequency-dependent reflectivity \cite{Morgan2007}. There is a slight frequency difference in the maximum transmission for the ``toward'' configuration compared to the minimum in the ``away'' configuration, attributed to the impedance mismatching of the transducers to the system \SI{50}{\ohm} impedance.

Fitting the detailed model to the data, we can extract the free velocity $v_\mathrm{f}=\SI{3865}{\meter\per\second}$, \SI{3}{\percent} slower than in the literature \cite{Morgan2007}, and the piezoelectric coupling $\Delta v/v = \SI{1.9}{\percent}$, where for a similar two-finger SPUDT design, but at a lower center frequency, a comparison value \cite{Martin2011} is  $\Delta v/v = \SI{1.5}{\percent}$. The shape of the overall response gives the imaginary reflectivity $r_\mathrm{s} = i$~$0.04$, which is roughly \SI{30}{\percent} higher than the RAM estimate for this geometry. We note that the positive sign for the reflectivity is expected for this type of substrate, and is responsible for inverting the preferred emission direction for this DART design compared to other substrate choices \cite{Morgan2007}.
The fit geometric DART capacitance of \SI{1.43}{\pico \farad} is \SI{22}{\percent} smaller than the RAM estimate. While this represents a sizeable discrepancy, this is consistent with other experiments we have completed with significantly different designs but at similar frequencies on similar substrates\cite{Satzinger2018}.

The rapid oscillations in the transmission signal near the operating frequency are due to interference of signals in the delay line, and give us a very precise way to measure the effective distance between the two transducers, which is \SI{400.08}{\micro\meter}, \SI{0.02}{\percent} longer than geometric center-to-center design distance. The position of the minimum dip in the ``away'' configuration is determined by the distance between the transduction (TC) and reflection (RC) centers, $3 \lambda_0/8$ by design, and is used as a fit parameter in the RAM theory. The numerical comparison yields a RC-TC distance \SI{4}{\percent} larger than the design value.

\begin{figure}[htbp]
\centering
\includegraphics[width=0.99\linewidth]{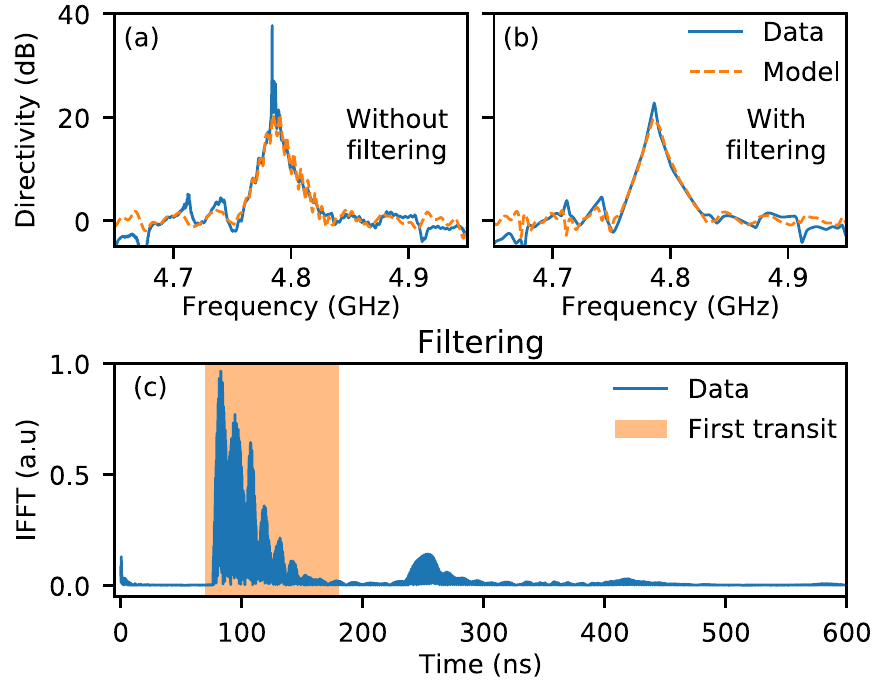}
\caption{DART directivity $D$, see Eq.~(\ref{eq.direct}), from data in Fig.~\ref{fig-2}. Solid and dashed lines represent the data and model, respectively.
(a) Unfiltered data.
(b) Data set filtered in time as per panel (c), representing the first transit peaking starting at $\cong 77$ ns, and the third transit starting at $\cong 235$ ns. Data are selected from the first transit, from \SI{70}{\nano\second} to \SI{180}{\nano\second} (colored band), and used to calculate the directivity in (b).}
\label{fig-3}
\end{figure}

The DART directivity is defined\cite{Morgan2007} as the ratio of the relevant $P$-matrix couplings, $D = |P_{13}/P_{23}|$. As these are not experimentally accessible, we instead use the approximate expression\cite{directivity}
\begin{equation}\label{eq.direct}
   D \approx \sqrt{\left | \frac{S_{21,t}}{S_{21,a}} \right |},
\end{equation}
the square root of the ratio of the transmission scattering parameters in the ``towards'' ($t$) and ``away'' ($a$) configurations. The directivity is shown in Fig.~\ref{fig-3}, where in (a) we show the directivity from the full frequency-dependent signal, and in (b) we show the directivity when the delay line response is filtered in the time domain to isolate the response from the first transit signal through the delay line, as shown in (c). We note that without filtering, the apparent peak directivity is as high as \SI{37}{\decibel}, but by selecting only the signal associated with the first transit, we find a more accurate measure of the peak directivity of \SI{23}{\decibel}.
This device has a bandwidth of \SI{15}{\mega\hertz}, defined as the frequency range over which the directivity exceeds \SI{15}{\decibel}.

We have measured a number of different DART configurations and tabulated representative results in Table~\ref{table:1}, using the first-transit filtered data for each design.
The highest directivity was measured to be $D \approx \SI{27}{\decibel}$ for an aperture $W = \SI{18}{\micro\meter}$, with $N=125$ cells.
The largest bandwidth was $\cong \SI{17}{\mega\hertz}$.
The directivity shows a weak dependence on DART aperture, with larger apertures giving slightly lower directivity and smaller bandwidth, with the change mostly due to an increase in the ``away" transmission.
These are, to our knowledge, the first GHz-frequency DART designs to yield directivities greater than \SI{25}{\decibel} with bandwidths approaching \SI{20}{\mega\hertz}.

\begin{table}[h]
\centering
\begin{tabular}{cccc}
\hline
\hline
$N$ & Aperture (\SI{}{\micro\meter}) & $D$  (dB) & BW  (MHz)\\
\hline
125 & 18 & 26.89 & 16.75 \\
    & 21 & 22.76 & 15.10 \\
    & 24 & 20.38 & 15.95 \\
\hline
130 & 18 & 25.89 & 16.00 \\
    & 21 & 23.21 & 15.85 \\
    & 24 & 19.36 & 15.20 \\
\hline
135 & 18 & 23.16 & 16.90 \\
    & 21 & 23.38 & 14.90 \\
    & 24 & 20.04 & 16.35 \\
\hline
\hline
\end{tabular}

\caption{Summary of different DART designs with their measured directivities and bandwidths. $N$ is the number of cells, $D$ the directivity and BW the bandwidth over which the directivity is greater than \SI{15}{\decibel}.}
\label{table:1}
\end{table}

In conclusion, we have fabricated DARTs with a high center frequency of \SI{4.8}{\giga\hertz}, demonstrating a directivity greater than \SI{15}{\decibel} over a bandwidth of \SI{17}{\mega\hertz}.
While highly promising, more study should be devoted to understanding the performance as a function of the transducer electrode dimensions.
Future work should study the dependence of the directivity and bandwidth on parameters such as the metal electrode thickness, as well as the electrode and inter-electrode width ratio, which here were kept close to unity.
The bandwidth of the DART could be increased by using slanted fingers \cite{Yatsuda1990}.
A weighted or resonant design\cite{Morgan2007,1994Ventura} could also sharpen and flatten the DART transduction peak response.
The performance demonstrated here is promising for experiments in the quantum limit.

Devices and experiments were supported by the Air Force Office of Scientific Research and the Army Research Laboratory, and material for this work was supported by the Department of Energy (DOE).
\'E. D. was supported by LDRD funds from Argonne National Laboratory, K.J.S. was supported by NSF GRFP (NSF DGE-1144085) and A.N.C. was supported by the DOE, Office of Basic Energy Sciences.
This work was partially supported by the UChicago MRSEC (NSF DMR-1420709) and made use of the Pritzker Nanofabrication Facility, which receives support from SHyNE, a node of the National Science Foundation’s National Nanotechnology Coordinated Infrastructure (NSF NNCI-1542205).

\bibliographystyle{apsrev4-1}
\bibliography{bib_main}

\begin{thebibliography}{21}%
\makeatletter
\providecommand \@ifxundefined [1]{%
 \@ifx{#1\undefined}
}%
\providecommand \@ifnum [1]{%
 \ifnum #1\expandafter \@firstoftwo
 \else \expandafter \@secondoftwo
 \fi
}%
\providecommand \@ifx [1]{%
 \ifx #1\expandafter \@firstoftwo
 \else \expandafter \@secondoftwo
 \fi
}%
\providecommand \natexlab [1]{#1}%
\providecommand \enquote  [1]{``#1''}%
\providecommand \bibnamefont  [1]{#1}%
\providecommand \bibfnamefont [1]{#1}%
\providecommand \citenamefont [1]{#1}%
\providecommand \href@noop [0]{\@secondoftwo}%
\providecommand \href [0]{\begingroup \@sanitize@url \@href}%
\providecommand \@href[1]{\@@startlink{#1}\@@href}%
\providecommand \@@href[1]{\endgroup#1\@@endlink}%
\providecommand \@sanitize@url [0]{\catcode `\\12\catcode `\$12\catcode
  `\&12\catcode `\#12\catcode `\^12\catcode `\_12\catcode `\%12\relax}%
\providecommand \@@startlink[1]{}%
\providecommand \@@endlink[0]{}%
\providecommand \url  [0]{\begingroup\@sanitize@url \@url }%
\providecommand \@url [1]{\endgroup\@href {#1}{\urlprefix }}%
\providecommand \urlprefix  [0]{URL }%
\providecommand \Eprint [0]{\href }%
\providecommand \doibase [0]{http://dx.doi.org/}%
\providecommand \selectlanguage [0]{\@gobble}%
\providecommand \bibinfo  [0]{\@secondoftwo}%
\providecommand \bibfield  [0]{\@secondoftwo}%
\providecommand \translation [1]{[#1]}%
\providecommand \BibitemOpen [0]{}%
\providecommand \bibitemStop [0]{}%
\providecommand \bibitemNoStop [0]{.\EOS\space}%
\providecommand \EOS [0]{\spacefactor3000\relax}%
\providecommand \BibitemShut  [1]{\csname bibitem#1\endcsname}%
\let\auto@bib@innerbib\@empty
\bibitem [{\citenamefont {O'Connell}\ \emph {et~al.}(2010)\citenamefont
  {O'Connell}, \citenamefont {Hofheinz}, \citenamefont {Ansmann}, \citenamefont
  {Bialczak}, \citenamefont {Lenander}, \citenamefont {Lucero}, \citenamefont
  {Neeley}, \citenamefont {Sank}, \citenamefont {Wang}, \citenamefont {Weides},
  \citenamefont {Wenner}, \citenamefont {Martinis},\ and\ \citenamefont
  {Cleland}}]{OConnell2010}%
  \BibitemOpen
  \bibfield  {author} {\bibinfo {author} {\bibfnamefont {A.~D.}\ \bibnamefont
  {O'Connell}}, \bibinfo {author} {\bibfnamefont {M.}~\bibnamefont {Hofheinz}},
  \bibinfo {author} {\bibfnamefont {M.}~\bibnamefont {Ansmann}}, \bibinfo
  {author} {\bibfnamefont {R.~C.}\ \bibnamefont {Bialczak}}, \bibinfo {author}
  {\bibfnamefont {M.}~\bibnamefont {Lenander}}, \bibinfo {author}
  {\bibfnamefont {E.}~\bibnamefont {Lucero}}, \bibinfo {author} {\bibfnamefont
  {M.}~\bibnamefont {Neeley}}, \bibinfo {author} {\bibfnamefont
  {D.}~\bibnamefont {Sank}}, \bibinfo {author} {\bibfnamefont {H.}~\bibnamefont
  {Wang}}, \bibinfo {author} {\bibfnamefont {M.}~\bibnamefont {Weides}},
  \bibinfo {author} {\bibfnamefont {J.}~\bibnamefont {Wenner}}, \bibinfo
  {author} {\bibfnamefont {J.~M.}\ \bibnamefont {Martinis}}, \ and\ \bibinfo
  {author} {\bibfnamefont {A.~N.}\ \bibnamefont {Cleland}},\ }\href {\doibase
  10.1038/nature08967} {\bibfield  {journal} {\bibinfo  {journal} {Nature}\
  }\textbf {\bibinfo {volume} {464}},\ \bibinfo {pages} {697} (\bibinfo {year}
  {2010})}\BibitemShut {NoStop}%
\bibitem [{\citenamefont {Chu}\ \emph {et~al.}(2017)\citenamefont {Chu},
  \citenamefont {Kharel}, \citenamefont {Renninger}, \citenamefont {Burkhart},
  \citenamefont {Frunzio}, \citenamefont {Rakich},\ and\ \citenamefont
  {Schoelkopf}}]{Chu2017}%
  \BibitemOpen
  \bibfield  {author} {\bibinfo {author} {\bibfnamefont {Y.}~\bibnamefont
  {Chu}}, \bibinfo {author} {\bibfnamefont {P.}~\bibnamefont {Kharel}},
  \bibinfo {author} {\bibfnamefont {W.~H.}\ \bibnamefont {Renninger}}, \bibinfo
  {author} {\bibfnamefont {L.~D.}\ \bibnamefont {Burkhart}}, \bibinfo {author}
  {\bibfnamefont {L.}~\bibnamefont {Frunzio}}, \bibinfo {author} {\bibfnamefont
  {P.~T.}\ \bibnamefont {Rakich}}, \ and\ \bibinfo {author} {\bibfnamefont
  {R.~J.}\ \bibnamefont {Schoelkopf}},\ }\href {\doibase
  10.1126/science.aao1511} {\bibfield  {journal} {\bibinfo  {journal}
  {Science}\ }\textbf {\bibinfo {volume} {358}},\ \bibinfo {pages} {199}
  (\bibinfo {year} {2017})}\BibitemShut {NoStop}%
\bibitem [{\citenamefont {Satzinger}\ \emph {et~al.}(2018)\citenamefont
  {Satzinger}, \citenamefont {Zhong}, \citenamefont {Chang}, \citenamefont
  {Peairs}, \citenamefont {Bienfait}, \citenamefont {Chou}, \citenamefont
  {Cleland}, \citenamefont {Conner}, \citenamefont {Dumur}, \citenamefont
  {Grebel}, \citenamefont {Gutierrez}, \citenamefont {November}, \citenamefont
  {Povey}, \citenamefont {Whiteley}, \citenamefont {Awschalom}, \citenamefont
  {Schuster},\ and\ \citenamefont {Cleland}}]{Satzinger2018}%
  \BibitemOpen
  \bibfield  {author} {\bibinfo {author} {\bibfnamefont {K.~J.}\ \bibnamefont
  {Satzinger}}, \bibinfo {author} {\bibfnamefont {Y.~P.}\ \bibnamefont
  {Zhong}}, \bibinfo {author} {\bibfnamefont {H.-S.}\ \bibnamefont {Chang}},
  \bibinfo {author} {\bibfnamefont {G.~A.}\ \bibnamefont {Peairs}}, \bibinfo
  {author} {\bibfnamefont {A.}~\bibnamefont {Bienfait}}, \bibinfo {author}
  {\bibfnamefont {M.-H.}\ \bibnamefont {Chou}}, \bibinfo {author}
  {\bibfnamefont {A.~Y.}\ \bibnamefont {Cleland}}, \bibinfo {author}
  {\bibfnamefont {C.~R.}\ \bibnamefont {Conner}}, \bibinfo {author}
  {\bibfnamefont {Ã.}~\bibnamefont {Dumur}}, \bibinfo {author} {\bibfnamefont
  {J.}~\bibnamefont {Grebel}}, \bibinfo {author} {\bibfnamefont
  {I.}~\bibnamefont {Gutierrez}}, \bibinfo {author} {\bibfnamefont {B.~H.}\
  \bibnamefont {November}}, \bibinfo {author} {\bibfnamefont {R.~G.}\
  \bibnamefont {Povey}}, \bibinfo {author} {\bibfnamefont {S.~J.}\ \bibnamefont
  {Whiteley}}, \bibinfo {author} {\bibfnamefont {D.~D.}\ \bibnamefont
  {Awschalom}}, \bibinfo {author} {\bibfnamefont {D.~I.}\ \bibnamefont
  {Schuster}}, \ and\ \bibinfo {author} {\bibfnamefont {A.~N.}\ \bibnamefont
  {Cleland}},\ }\href {\doibase s41586-018-0719-5} {\bibfield  {journal}
  {\bibinfo  {journal} {Nature}\ }\textbf {\bibinfo {volume} {563}},\ \bibinfo
  {pages} {661} (\bibinfo {year} {2018})}\BibitemShut {NoStop}%
\bibitem [{\citenamefont {Morgan}(2007)}]{Morgan2007}%
  \BibitemOpen
  \bibfield  {author} {\bibinfo {author} {\bibfnamefont {D.}~\bibnamefont
  {Morgan}},\ }\href@noop {} {\emph {\bibinfo {title} {Surface acoustic wave
  filters}}},\ \bibinfo {edition} {2nd}\ ed.\ (\bibinfo  {publisher} {Elsevier
  Ltd.},\ \bibinfo {year} {2007})\BibitemShut {NoStop}%
\bibitem [{\citenamefont {Gustafsson}\ \emph {et~al.}(2014)\citenamefont
  {Gustafsson}, \citenamefont {Aref}, \citenamefont {Kockum}, \citenamefont
  {Ekstr{\"o}m}, \citenamefont {Johansson},\ and\ \citenamefont
  {Delsing}}]{Gustafsson2014}%
  \BibitemOpen
  \bibfield  {author} {\bibinfo {author} {\bibfnamefont {M.~V.}\ \bibnamefont
  {Gustafsson}}, \bibinfo {author} {\bibfnamefont {T.}~\bibnamefont {Aref}},
  \bibinfo {author} {\bibfnamefont {A.~F.}\ \bibnamefont {Kockum}}, \bibinfo
  {author} {\bibfnamefont {M.~K.}\ \bibnamefont {Ekstr{\"o}m}}, \bibinfo
  {author} {\bibfnamefont {G.}~\bibnamefont {Johansson}}, \ and\ \bibinfo
  {author} {\bibfnamefont {P.}~\bibnamefont {Delsing}},\ }\href {\doibase
  10.1126/science.1257219} {\bibfield  {journal} {\bibinfo  {journal}
  {Science}\ }\textbf {\bibinfo {volume} {346}},\ \bibinfo {pages} {207}
  (\bibinfo {year} {2014})}\BibitemShut {NoStop}%
\bibitem [{\citenamefont {Manenti}\ \emph {et~al.}(2017)\citenamefont
  {Manenti}, \citenamefont {Kockum}, \citenamefont {Patterson}, \citenamefont
  {Behrle}, \citenamefont {Rahamim}, \citenamefont {Tancredi}, \citenamefont
  {Nori},\ and\ \citenamefont {Leek}}]{Manenti2017}%
  \BibitemOpen
  \bibfield  {author} {\bibinfo {author} {\bibfnamefont {R.}~\bibnamefont
  {Manenti}}, \bibinfo {author} {\bibfnamefont {A.~F.}\ \bibnamefont {Kockum}},
  \bibinfo {author} {\bibfnamefont {A.}~\bibnamefont {Patterson}}, \bibinfo
  {author} {\bibfnamefont {T.}~\bibnamefont {Behrle}}, \bibinfo {author}
  {\bibfnamefont {J.}~\bibnamefont {Rahamim}}, \bibinfo {author} {\bibfnamefont
  {G.}~\bibnamefont {Tancredi}}, \bibinfo {author} {\bibfnamefont
  {F.}~\bibnamefont {Nori}}, \ and\ \bibinfo {author} {\bibfnamefont {P.~J.}\
  \bibnamefont {Leek}},\ }\href {\doibase 10.1038/s41467-017-01063-9}
  {\bibfield  {journal} {\bibinfo  {journal} {Nature Communications}\ }\textbf
  {\bibinfo {volume} {8}},\ \bibinfo {pages} {975} (\bibinfo {year}
  {2017})}\BibitemShut {NoStop}%
\bibitem [{\citenamefont {Moores}\ \emph {et~al.}(2018)\citenamefont {Moores},
  \citenamefont {Sletten}, \citenamefont {Viennot},\ and\ \citenamefont
  {Lehnert}}]{Moores2017}%
  \BibitemOpen
  \bibfield  {author} {\bibinfo {author} {\bibfnamefont {B.~A.}\ \bibnamefont
  {Moores}}, \bibinfo {author} {\bibfnamefont {L.~R.}\ \bibnamefont {Sletten}},
  \bibinfo {author} {\bibfnamefont {J.~J.}\ \bibnamefont {Viennot}}, \ and\
  \bibinfo {author} {\bibfnamefont {K.~W.}\ \bibnamefont {Lehnert}},\ }\href
  {\doibase 10.1103/PhysRevLett.120.227701} {\bibfield  {journal} {\bibinfo
  {journal} {Phys. Rev. Lett.}\ }\textbf {\bibinfo {volume} {120}},\ \bibinfo
  {pages} {227701} (\bibinfo {year} {2018})}\BibitemShut {NoStop}%
\bibitem [{\citenamefont {Hanma}\ and\ \citenamefont
  {Hunsinger}(1976)}]{Hanma1976}%
  \BibitemOpen
  \bibfield  {author} {\bibinfo {author} {\bibfnamefont {K.}~\bibnamefont
  {Hanma}}\ and\ \bibinfo {author} {\bibfnamefont {B.~J.}\ \bibnamefont
  {Hunsinger}},\ }in\ \href {\doibase 10.1109/ULTSYM.1976.196692} {\emph
  {\bibinfo {booktitle} {1976 Ultrasonics Symposium}}}\ (\bibinfo {year}
  {1976})\ pp.\ \bibinfo {pages} {328--331}\BibitemShut {NoStop}%
\bibitem [{\citenamefont {Hartmann}\ \emph {et~al.}(1982)\citenamefont
  {Hartmann}, \citenamefont {Wright}, \citenamefont {Kansy},\ and\
  \citenamefont {Garber}}]{Hartmann1982}%
  \BibitemOpen
  \bibfield  {author} {\bibinfo {author} {\bibfnamefont {C.~S.}\ \bibnamefont
  {Hartmann}}, \bibinfo {author} {\bibfnamefont {P.~V.}\ \bibnamefont
  {Wright}}, \bibinfo {author} {\bibfnamefont {R.~J.}\ \bibnamefont {Kansy}}, \
  and\ \bibinfo {author} {\bibfnamefont {E.~M.}\ \bibnamefont {Garber}},\ }in\
  \href {\doibase 10.1109/ULTSYM.1982.197784} {\emph {\bibinfo {booktitle}
  {1982 Ultrasonics Symposium}}}\ (\bibinfo {year} {1982})\ pp.\ \bibinfo
  {pages} {40--45}\BibitemShut {NoStop}%
\bibitem [{\citenamefont {Lewis}(1983)}]{Lewis1983}%
  \BibitemOpen
  \bibfield  {author} {\bibinfo {author} {\bibfnamefont {M.}~\bibnamefont
  {Lewis}},\ }in\ \href {\doibase 10.1109/ULTSYM.1983.198024} {\emph {\bibinfo
  {booktitle} {1983 Ultrasonics Symposium}}}\ (\bibinfo {year} {1983})\ pp.\
  \bibinfo {pages} {104--108}\BibitemShut {NoStop}%
\bibitem [{\citenamefont {Kodama}\ \emph {et~al.}(1986)\citenamefont {Kodama},
  \citenamefont {Kawabata}, \citenamefont {Yasuhara},\ and\ \citenamefont
  {Sato}}]{Kodama1986}%
  \BibitemOpen
  \bibfield  {author} {\bibinfo {author} {\bibfnamefont {T.}~\bibnamefont
  {Kodama}}, \bibinfo {author} {\bibfnamefont {H.}~\bibnamefont {Kawabata}},
  \bibinfo {author} {\bibfnamefont {Y.}~\bibnamefont {Yasuhara}}, \ and\
  \bibinfo {author} {\bibfnamefont {H.}~\bibnamefont {Sato}},\ }in\ \href
  {\doibase 10.1109/ULTSYM.1986.198710} {\emph {\bibinfo {booktitle} {IEEE 1986
  Ultrasonics Symposium}}}\ (\bibinfo {year} {1986})\ pp.\ \bibinfo {pages}
  {59--64}\BibitemShut {NoStop}%
\bibitem [{\citenamefont {Hode}\ \emph {et~al.}(1990)\citenamefont {Hode},
  \citenamefont {Doisy},\ and\ \citenamefont {Dufilie}}]{Hode1990}%
  \BibitemOpen
  \bibfield  {author} {\bibinfo {author} {\bibfnamefont {J.~M.}\ \bibnamefont
  {Hode}}, \bibinfo {author} {\bibfnamefont {M.}~\bibnamefont {Doisy}}, \ and\
  \bibinfo {author} {\bibfnamefont {P.}~\bibnamefont {Dufilie}},\ }in\ \href
  {\doibase 10.1109/ULTSYM.1990.171402} {\emph {\bibinfo {booktitle} {IEEE
  Symposium on Ultrasonics}}},\ Vol.~\bibinfo {volume} {1}\ (\bibinfo {year}
  {1990})\ pp.\ \bibinfo {pages} {429--434}\BibitemShut {NoStop}%
\bibitem [{\citenamefont {Yamanouchi}\ \emph {et~al.}(1992)\citenamefont
  {Yamanouchi}, \citenamefont {Lee}, \citenamefont {Yamamoto}, \citenamefont
  {Meguro},\ and\ \citenamefont {Odagawa}}]{Yamanouchi1992}%
  \BibitemOpen
  \bibfield  {author} {\bibinfo {author} {\bibfnamefont {K.}~\bibnamefont
  {Yamanouchi}}, \bibinfo {author} {\bibfnamefont {C.~H.~S.}\ \bibnamefont
  {Lee}}, \bibinfo {author} {\bibfnamefont {K.}~\bibnamefont {Yamamoto}},
  \bibinfo {author} {\bibfnamefont {T.}~\bibnamefont {Meguro}}, \ and\ \bibinfo
  {author} {\bibfnamefont {H.}~\bibnamefont {Odagawa}},\ }in\ \href {\doibase
  10.1109/ULTSYM.1992.276049} {\emph {\bibinfo {booktitle} {IEEE 1992
  Ultrasonics Symposium Proceedings}}},\ Vol.~\bibinfo {volume} {1}\ (\bibinfo
  {year} {1992})\ pp.\ \bibinfo {pages} {139--142}\BibitemShut {NoStop}%
\bibitem [{\citenamefont {Garber}\ \emph {et~al.}(1994)\citenamefont {Garber},
  \citenamefont {Yip},\ and\ \citenamefont {Henderson}}]{Garber1994}%
  \BibitemOpen
  \bibfield  {author} {\bibinfo {author} {\bibfnamefont {E.~M.}\ \bibnamefont
  {Garber}}, \bibinfo {author} {\bibfnamefont {D.~S.}\ \bibnamefont {Yip}}, \
  and\ \bibinfo {author} {\bibfnamefont {D.~K.}\ \bibnamefont {Henderson}},\
  }\href {\doibase 10.1109/ULTSYM.1994.401543} {\bibfield  {journal} {\bibinfo
  {journal} {Proceedings of {IEEE} Ultrasonics Symposium}\ }\textbf {\bibinfo
  {volume} {1}},\ \bibinfo {pages} {7} (\bibinfo {year} {1994})}\BibitemShut
  {NoStop}%
\bibitem [{\citenamefont {Lehtonen}\ \emph {et~al.}(2003)\citenamefont
  {Lehtonen}, \citenamefont {Plessky}, \citenamefont {Hartmann},\ and\
  \citenamefont {Salomaa}}]{Lehtonen2003}%
  \BibitemOpen
  \bibfield  {author} {\bibinfo {author} {\bibfnamefont {S.}~\bibnamefont
  {Lehtonen}}, \bibinfo {author} {\bibfnamefont {V.~P.}\ \bibnamefont
  {Plessky}}, \bibinfo {author} {\bibfnamefont {C.~S.}\ \bibnamefont
  {Hartmann}}, \ and\ \bibinfo {author} {\bibfnamefont {M.~M.}\ \bibnamefont
  {Salomaa}},\ }\href {\doibase 10.1109/TUFFC.2003.1251122} {\bibfield
  {journal} {\bibinfo  {journal} {IEEE Transactions on Ultrasonics,
  Ferroelectrics, and Frequency Control}\ }\textbf {\bibinfo {volume} {50}},\
  \bibinfo {pages} {1404} (\bibinfo {year} {2003})}\BibitemShut {NoStop}%
\bibitem [{\citenamefont {Martin}\ \emph {et~al.}(2011)\citenamefont {Martin},
  \citenamefont {Biryukov}, \citenamefont {Schmidt}, \citenamefont {Steiner},\
  and\ \citenamefont {Wall}}]{Martin2011}%
  \BibitemOpen
  \bibfield  {author} {\bibinfo {author} {\bibfnamefont {G.}~\bibnamefont
  {Martin}}, \bibinfo {author} {\bibfnamefont {S.~V.}\ \bibnamefont
  {Biryukov}}, \bibinfo {author} {\bibfnamefont {H.}~\bibnamefont {Schmidt}},
  \bibinfo {author} {\bibfnamefont {B.}~\bibnamefont {Steiner}}, \ and\
  \bibinfo {author} {\bibfnamefont {B.}~\bibnamefont {Wall}},\ }\href {\doibase
  10.1109/TUFFC.2011.1849} {\bibfield  {journal} {\bibinfo  {journal} {{IEEE}
  Transactions on Ultrasonics, Ferroelectrics, and Frequency Control}\ }\textbf
  {\bibinfo {volume} {58}},\ \bibinfo {pages} {658} (\bibinfo {year}
  {2011})}\BibitemShut {NoStop}%
\bibitem [{\citenamefont {Dufilie}\ and\ \citenamefont
  {Ventura}(1995)}]{Dufilie1995}%
  \BibitemOpen
  \bibfield  {author} {\bibinfo {author} {\bibfnamefont {P.}~\bibnamefont
  {Dufilie}}\ and\ \bibinfo {author} {\bibfnamefont {P.}~\bibnamefont
  {Ventura}},\ }in\ \href {\doibase 10.1109/ULTSYM.1995.495532} {\emph
  {\bibinfo {booktitle} {IEEE Ultrasonics Symposium}}},\ Vol.~\bibinfo {volume}
  {1}\ (\bibinfo {year} {1995})\ pp.\ \bibinfo {pages} {13--16}\BibitemShut
  {NoStop}%
\bibitem [{\citenamefont {Morgan}(1998)}]{Morgan1998}%
  \BibitemOpen
  \bibfield  {author} {\bibinfo {author} {\bibfnamefont {D.~P.}\ \bibnamefont
  {Morgan}},\ }\href {\doibase 10.1109/58.646919} {\bibfield  {journal}
  {\bibinfo  {journal} {IEEE Transactions on Ultrasonics, Ferroelectrics, and
  Frequency Control}\ }\textbf {\bibinfo {volume} {45}},\ \bibinfo {pages}
  {152} (\bibinfo {year} {1998})}\BibitemShut {NoStop}%
\bibitem [{dir()}]{directivity}%
  \BibitemOpen
  \href@noop {} {\bibinfo  {journal} {See supplementary material for derivation
  and discussion}\ }\BibitemShut {NoStop}%
\bibitem [{\citenamefont {Yatsuda}\ \emph {et~al.}(1990)\citenamefont
  {Yatsuda}, \citenamefont {Takeuchi},\ and\ \citenamefont
  {Yoshikawa}}]{Yatsuda1990}%
  \BibitemOpen
\bibfield  {journal} {  }\bibfield  {author} {\bibinfo {author} {\bibfnamefont
  {H.}~\bibnamefont {Yatsuda}}, \bibinfo {author} {\bibfnamefont
  {Y.}~\bibnamefont {Takeuchi}}, \ and\ \bibinfo {author} {\bibfnamefont
  {S.}~\bibnamefont {Yoshikawa}},\ }in\ \href {\doibase
  10.1109/ULTSYM.1990.171327} {\emph {\bibinfo {booktitle} {IEEE Symposium on
  Ultrasonics}}},\ Vol.~\bibinfo {volume} {1}\ (\bibinfo {year} {1990})\ pp.\
  \bibinfo {pages} {61--66}\BibitemShut {NoStop}%
\bibitem [{\citenamefont {Ventura}\ \emph {et~al.}(1994)\citenamefont
  {Ventura}, \citenamefont {Solal}, \citenamefont {Dufilie}, \citenamefont
  {Hode},\ and\ \citenamefont {Roux}}]{1994Ventura}%
  \BibitemOpen
  \bibfield  {author} {\bibinfo {author} {\bibfnamefont {P.}~\bibnamefont
  {Ventura}}, \bibinfo {author} {\bibfnamefont {M.}~\bibnamefont {Solal}},
  \bibinfo {author} {\bibfnamefont {P.}~\bibnamefont {Dufilie}}, \bibinfo
  {author} {\bibfnamefont {J.~M.}\ \bibnamefont {Hode}}, \ and\ \bibinfo
  {author} {\bibfnamefont {F.}~\bibnamefont {Roux}},\ }in\ \href {\doibase
  10.1109/ULTSYM.1994.401542} {\emph {\bibinfo {booktitle} {1994 Proceedings of
  IEEE Ultrasonics Symposium}}},\ Vol.~\bibinfo {volume} {1}\ (\bibinfo {year}
  {1994})\ pp.\ \bibinfo {pages} {1--6 vol.1}\BibitemShut {NoStop}%
\end{thebibliography}%


\begin{thebibliography}{3}%
\makeatletter
\providecommand \@ifxundefined [1]{%
 \@ifx{#1\undefined}
}%
\providecommand \@ifnum [1]{%
 \ifnum #1\expandafter \@firstoftwo
 \else \expandafter \@secondoftwo
 \fi
}%
\providecommand \@ifx [1]{%
 \ifx #1\expandafter \@firstoftwo
 \else \expandafter \@secondoftwo
 \fi
}%
\providecommand \natexlab [1]{#1}%
\providecommand \enquote  [1]{``#1''}%
\providecommand \bibnamefont  [1]{#1}%
\providecommand \bibfnamefont [1]{#1}%
\providecommand \citenamefont [1]{#1}%
\providecommand \href@noop [0]{\@secondoftwo}%
\providecommand \href [0]{\begingroup \@sanitize@url \@href}%
\providecommand \@href[1]{\@@startlink{#1}\@@href}%
\providecommand \@@href[1]{\endgroup#1\@@endlink}%
\providecommand \@sanitize@url [0]{\catcode `\\12\catcode `\$12\catcode
  `\&12\catcode `\#12\catcode `\^12\catcode `\_12\catcode `\%12\relax}%
\providecommand \@@startlink[1]{}%
\providecommand \@@endlink[0]{}%
\providecommand \url  [0]{\begingroup\@sanitize@url \@url }%
\providecommand \@url [1]{\endgroup\@href {#1}{\urlprefix }}%
\providecommand \urlprefix  [0]{URL }%
\providecommand \Eprint [0]{\href }%
\providecommand \doibase [0]{http://dx.doi.org/}%
\providecommand \selectlanguage [0]{\@gobble}%
\providecommand \bibinfo  [0]{\@secondoftwo}%
\providecommand \bibfield  [0]{\@secondoftwo}%
\providecommand \translation [1]{[#1]}%
\providecommand \BibitemOpen [0]{}%
\providecommand \bibitemStop [0]{}%
\providecommand \bibitemNoStop [0]{.\EOS\space}%
\providecommand \EOS [0]{\spacefactor3000\relax}%
\providecommand \BibitemShut  [1]{\csname bibitem#1\endcsname}%
\let\auto@bib@innerbib\@empty
\bibitem [{\citenamefont {Morgan}(2007)}]{2007Morg}%
  \BibitemOpen
  \bibfield  {author} {\bibinfo {author} {\bibfnamefont {D.}~\bibnamefont
  {Morgan}},\ }\href@noop {} {\emph {\bibinfo {title} {Surface Acoustic Wave
  Filters}}}\ (\bibinfo  {publisher} {Elsevier Ltd.},\ \bibinfo {year}
  {2007})\BibitemShut {NoStop}%
\bibitem [{\citenamefont {Tobolka}(1979)}]{Tobolka1979}%
  \BibitemOpen
  \bibfield  {author} {\bibinfo {author} {\bibfnamefont {G.}~\bibnamefont
  {Tobolka}},\ }\href {\doibase 10.1109/T-SU.1979.31128} {\bibfield  {journal}
  {\bibinfo  {journal} {IEEE Transactions on Sonics and Ultrasonics}\ }\textbf
  {\bibinfo {volume} {26}},\ \bibinfo {pages} {426} (\bibinfo {year}
  {1979})}\BibitemShut {NoStop}%
\bibitem [{\citenamefont {Pozar}(2012)}]{2009Pozar}%
  \BibitemOpen
  \bibfield  {author} {\bibinfo {author} {\bibfnamefont {D.~M.}\ \bibnamefont
  {Pozar}},\ }\href@noop {} {\emph {\bibinfo {title} {Microwave
  engineering}}},\ \bibinfo {edition} {4th}\ ed.\ (\bibinfo  {publisher} {John
  Wiley \& Sons},\ \bibinfo {year} {2012})\BibitemShut {NoStop}%
\end{thebibliography}%

\end{document}


\title{Supplementary Material for ``Unidirectional Distributed Acoustic Reflection Transducers for Quantum Applications"}

\author{\'E. Dumur}
\affiliation{Institute for Molecular Engineering, University of Chicago, Chicago IL 60637, USA}
\affiliation{Institute for Molecular Engineering and Materials Science Division, Argonne National Laboratory, Argonne, IL 60439, USA}
\author{K. J. Satzinger}
\altaffiliation[Present address: ]{Google, Santa Barbara, California 93117, USA}
\affiliation{Department of Physics, University of California, Santa Barbara CA 93106, USA}
\affiliation{Institute for Molecular Engineering, University of Chicago, Chicago IL 60637, USA}
\author{G. A. Peairs}
\affiliation{Department of Physics, University of California, Santa Barbara CA 93106, USA}
\affiliation{Institute for Molecular Engineering, University of Chicago, Chicago IL 60637, USA}
\author{Ming-Han Chou}
\affiliation{Institute for Molecular Engineering, University of Chicago, Chicago IL 60637, USA}
\affiliation{Department of Physics, University of Chicago, Chicago IL 60637, USA}
\author{A. Bienfait}
\affiliation{Institute for Molecular Engineering, University of Chicago, Chicago IL 60637, USA}
\author{H.-S. Chang}
\affiliation{Institute for Molecular Engineering, University of Chicago, Chicago IL 60637, USA}
\author{C. R. Conner}
\affiliation{Institute for Molecular Engineering, University of Chicago, Chicago IL 60637, USA}
\author{J. Grebel}
\affiliation{Institute for Molecular Engineering, University of Chicago, Chicago IL 60637, USA}
\author{R. G. Povey}
\affiliation{Institute for Molecular Engineering, University of Chicago, Chicago IL 60637, USA}
\affiliation{Department of Physics, University of Chicago, Chicago IL 60637, USA}
\author{Y. P. Zhong}
\affiliation{Institute for Molecular Engineering, University of Chicago, Chicago IL 60637, USA}
\author{A. N. Cleland}
\affiliation{Institute for Molecular Engineering, University of Chicago, Chicago IL 60637, USA}
\affiliation{Institute for Molecular Engineering and Materials Science Division, Argonne National Laboratory, Argonne, IL 60439, USA}

\begin{abstract}
We provide here a detailed derivation of the approximate directivity formula, Eq.~(1).
\end{abstract}

\maketitle

\section{From P-matrix to Y-matrix}

\begin{figure}[htbp]
\centering
\includegraphics[width=0.8\linewidth]{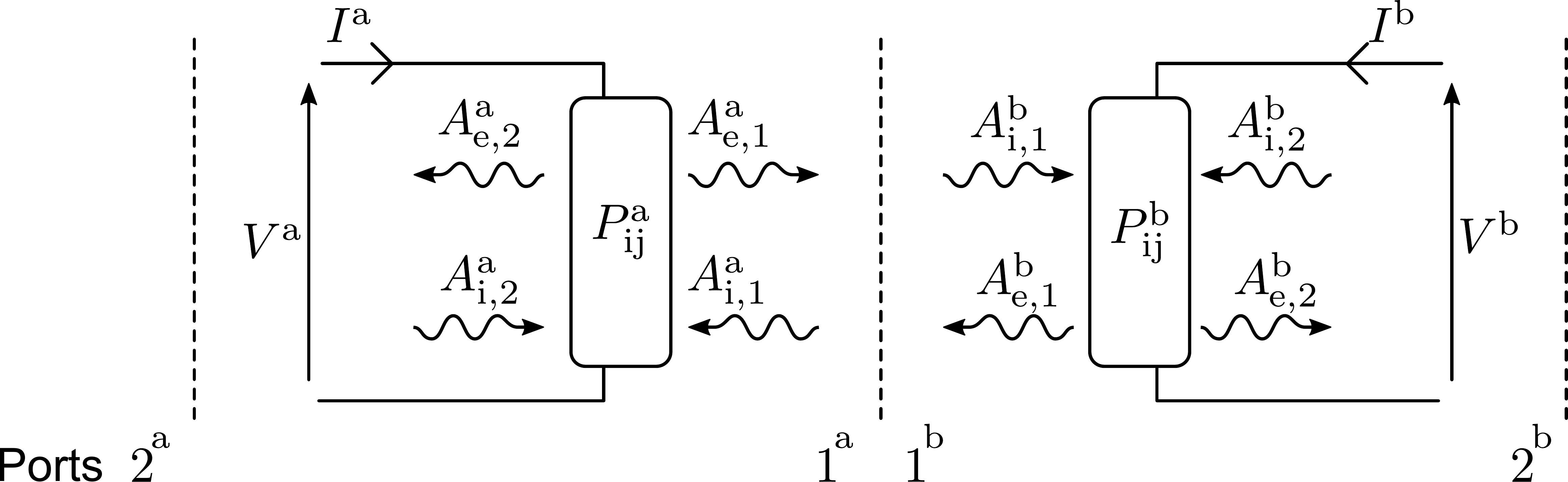}
\caption{Circuit model for two transducers $a$ and $b$, showing both the acoustic and electrical parameters. Ports 1 for each transducer are on their facing sides, while ports 2 face the exterior.}
\label{fig:circuit}
\end{figure}

We first introduce the $P$-matrix description of a surface acoustic wave (SAW) delay line, formed by two SAW transducers $a$ and $b$ spaced a distance $L$ apart, as in Ref. \onlinecite{2007Morg} and shown in Fig.~\ref{fig:circuit}. The $P$-matrix description for the transducers, as described in Ref. \onlinecite{Tobolka1979}, relates the electrical currents $I$ and voltages $V$ applied to each transducer $a$ and $b$ to the incident and emitted acoustic wave amplitudes $A_{i,e}$:
\begin{eqnarray}
    I^\mathrm{a} & = & P_{31}^\mathrm{a} A_{\mathrm{i}1}^\mathrm{a} + P_{32}^\mathrm{a} A_{\mathrm{i}2}^\mathrm{a} + P_{33}^\mathrm{a} V^\mathrm{a} \\
    I^\mathrm{b} & = & P_{31}^\mathrm{b} A_{\mathrm{i}1}^\mathrm{b} + P_{32}^\mathrm{b} A_{\mathrm{i}2}^\mathrm{b} + P_{33}^\mathrm{b} V^\mathrm{b}.
\end{eqnarray}
The amplitudes of the emitted SAWs are given by
\begin{eqnarray}
    A_{\mathrm{e}1}^\mathrm{a} & = & P_{11}^\mathrm{a} A_{\mathrm{i}1}^\mathrm{a} + P_{12}^\mathrm{a} A_{\mathrm{i}2}^\mathrm{a} + P_{13}^\mathrm{a} V^\mathrm{a} \\
    A_{\mathrm{e}1}^\mathrm{b} & = & P_{11}^\mathrm{b} A_{\mathrm{i}1}^\mathrm{b} + P_{12}^\mathrm{b} A_{\mathrm{i}2}^\mathrm{b} + P_{13}^\mathrm{b} V^\mathrm{b}.
\end{eqnarray}
We assume no acoustic signals are incident from outside the delay line structure, which translates to setting
\begin{equation}
    A_{\mathrm{i}2}^\mathrm{a} = A_{\mathrm{i}2}^\mathrm{b} = 0.
\end{equation}
This assumption can be inaccurate for delay lines that have SAWs that pass through the transducers and reflect off the edges of the device substrate. Here, experimental tests with and without absorbing pads outside the delay line structure show no noticeable difference in device performance, supporting this approximation.

The previous system of equations then simplifies to
\begin{eqnarray}\label{eq:0}
    I^\mathrm{a} & = & P_{31}^\mathrm{a} A_{\mathrm{i}1}^\mathrm{a} + P_{33}^\mathrm{a} V^\mathrm{a}, \\
    I^\mathrm{b} & = & P_{31}^\mathrm{b} A_{\mathrm{i}1}^\mathrm{b} + P_{33}^\mathrm{b} V^\mathrm{b},
\end{eqnarray}
and
\begin{eqnarray}\label{eq:1}
    A_{\mathrm{e}1}^\mathrm{a} & = &P_{11}^\mathrm{a} A_{\mathrm{i}1}^\mathrm{a} + P_{13}^\mathrm{a} V^\mathrm{a}, \\
    A_{\mathrm{e}1}^\mathrm{b} & = &P_{11}^\mathrm{b} A_{\mathrm{i}1}^\mathrm{b} + P_{13}^\mathrm{b} V^\mathrm{b}.
\end{eqnarray}
We assume that ports $1^\mathrm{a}$ and $1^\mathrm{b}$ are facing one another, as in Fig.~\ref{fig:circuit}, so that the transmitted and incident acoustic amplitudes are related by
\begin{eqnarray}\label{eq:2}
    A_{\mathrm{e}1}^\mathrm{a} &  = & A_{\mathrm{i}1}^\mathrm{b}, \\
    A_{\mathrm{e}1}^\mathrm{b} &  = & A_{\mathrm{i}1}^\mathrm{a}.
\end{eqnarray}

We can then solve for $A_{\mathrm{i}1}^\mathrm{a,b}$:
\begin{eqnarray}\label{eq:3}
    A_{\mathrm{i}1}^\mathrm{b} & = & \frac{P_{11}^\mathrm{a} P_{13}^\mathrm{b}}{1 - P_{11}^\mathrm{a} P_{11}^\mathrm{b} } V^\mathrm{b} + \frac{P_{13}^\mathrm{a}}{1 - P_{11}^\mathrm{a} P_{11}^\mathrm{b} } V^\mathrm{a} \\
    A_{\mathrm{i}1}^\mathrm{a} & = & \frac{P_{11}^\mathrm{b} P_{13}^\mathrm{a}}{1 - P_{11}^\mathrm{a} P_{11}^\mathrm{b} } V^\mathrm{a} + \frac{P_{13}^\mathrm{b}}{1 - P_{11}^\mathrm{a} P_{11}^\mathrm{b} } V^\mathrm{b}.
\end{eqnarray}
Solving for $V^\mathrm{a,b}$, we obtain the following admittance matrix directly relating the currents to the voltages,
\begin{equation}\label{eq:4}
    \left [ \begin{array}{c} I^\mathrm{a} \\ I^\mathrm{b} \end{array} \right ] = \left [ \begin{array}{cc} Y_{11} & Y_{12} \\ Y_{21} & Y_{22} \end{array} \right ] \left [ \begin{array}{c} V^\mathrm{a} \\ V^\mathrm{b} \end{array} \right ],
\end{equation}
with the admittance coefficients
\begin{align}
    Y_{11} & =  P_{33}^\mathrm{a} - \frac{2 P_{11}^\mathrm{b} \left( P_{13}^\mathrm{a} \right)^2}{1 - P_{11}^\mathrm{a} P_{11}^\mathrm{b}} \nonumber \\
    Y_{12} & =   Y_{21} = -\frac{2 P_{13}^\mathrm{a} P_{13}^\mathrm{b}}{1 - P_{11}^\mathrm{a} P_{11}^\mathrm{b}} \nonumber \\
    Y_{22} & =  P_{33}^\mathrm{b} - \frac{2 P_{11}^\mathrm{a} \left( P_{13}^\mathrm{b} \right)^2}{1 - P_{11}^\mathrm{a} P_{11}^\mathrm{b}}.
\end{align}
This result corresponds to Ref.~\onlinecite{2007Morg}, Eq.~(D.18).

\section{From Y-matrix to S-matrix}
The scattering $S$-matrix can be derived from the $Y$-matrix assuming a reference admittance $Y_0$, using a standard transformation\cite{2009Pozar}. The transmission coefficient $S_{21}$ is then
\begin{equation}
    S_{21} =  \frac{-2 Y_{21} Y_0}{\left( Y_{11} + Y_{0} \right ) \left ( Y_{22} + Y_{0} \right ) - Y_{12} Y_{21}}.
\end{equation}

This allows us to write the transmission in terms of the $P$-matrix components:
\begin{equation}
    S_{21} = \frac{-4 P_{13}^\mathrm{a} P_{13}^\mathrm{b} Y_0}{\Delta Y},
\end{equation}
with denominator
\begin{align}
    \Delta Y &= 2 \left (P_{13}^\mathrm{a}\right )^2 \left (2 \left (P_{13}^\mathrm{b}\right )^2 + P_{11}^\mathrm{b} \left (P_{33}^\mathrm{b} + Y_0\right )\right ) + \\
    &+  \left (P_{33}^\mathrm{a} + Y_0\right ) \left (2 P_{11}^\mathrm{a} \left ( P_{13}^\mathrm{b}\right )^2 - P_{33}^\mathrm{b} - Y_0 + P_{11}^\mathrm{a} P_{11}^\mathrm{b} \left (P_{33}^\mathrm{b} + Y_0\right )\right ).
\end{align}

\section{Transmission for the ``away" and ``toward" delay line configurations}

In order to derive a simpler expression for the transmission coefficient $S_{21}$, we assume that the $P$-matrices for the \textit{a} and \textit{b} transducers are identical, as applies to our transducer design. We also assume that the $P$ coefficients are determined solely by the transducers, ignoring any losses or phase delays associated with the distance $L$ separating the two transducers.  Any losses will affect the ``away'' and ``toward'' transmissions $S_{21}$ by the same amount, thus not affecting the directivity, which is the ratio of the transmissions. The phase delay due to the separation $L$ is only relevant to the directivity if interference effects between different delay line transits are allowed, but here we filter the data in the time domain to only include the first transit signal.

The ``toward'' and ``away'' transmission coefficients ($S_{21}^\mathrm{\rightarrow \leftarrow}$ and $S_{21}^\mathrm{\leftarrow \rightarrow}$ respectively) are then given by
\begin{align}
    S_{21}^\mathrm{\rightarrow \leftarrow} & = \frac{-4 \left (P_{13}\right )^2 Y_0}{\left(1-\left(P_{11}\right)^2\right)\left(\left( P_{33} + Y_0 -\frac{2 P_{11}\left(P_{13}\right)^2}{1-\left(P_{11}\right)^2}\right)^2 - \frac{4 P_{13}^4}{\left(1-\left(P_{11}\right)^2\right)^2}\right)}
\end{align}
and
\begin{align}
    S_{21}^\mathrm{\leftarrow \rightarrow} & = \frac{-4 \left(P_{23}\right)^2 Y_0}{\left(1-\left(P_{22}\right)^2\right)\left(\left( P_{33} + Y_0 -\frac{2 P_{22}\left(P_{23}\right)^2}{1-\left(P_{22}\right)^2}\right)^2 - \frac{4 P_{23}^4}{\left(1-\left(P_{11}\right)^2\right)^2}\right)},
\end{align}
where the ``away'' transmission is obtained by interchanging ports 1 and 2 in the ``toward'' expression.

\section{Directivity from transmission coefficient}

We next derive the ratio of the ``away'' and ``toward'' transmission as function of the $P$-matrix coefficients:
\begin{equation}\label{eq:initial-d}
    \frac{S_{21}^\mathrm{\rightarrow \leftarrow}}{S_{21}^\mathrm{\leftarrow \rightarrow}}  = \frac{\left(P_{{13}}\right)^2 \left(\left(P_{{11}}\right)^2 - 1\right) \left(4 \left(P_{{23}}\right)^4 - \left(2 P_{{22}} \left(P_{{23}}\right)^2 + \left(\left(P_{{22}}\right)^2 - 1\right) \left(P_{{33}} + Y_{0}\right)\right)^{2}\right)}{\left(P_{{23}}\right)^2 \left(4 \left(P_{{13}}\right)^4 - \left(2 P_{{11}} \left(P_{{13}}\right)^2 + \left(\left(P_{{11}}\right)^2 - 1\right) \left(P_{{33}} + Y_{0}\right)\right)^{2}\right) \left(\left(P_{{22}}\right)^2 - 1\right)}
\end{equation}
Now we look at the transmission ratio close to the delay line center frequency $f_0$, with frequency $f = f_0 + \Delta f$ where $\Delta f$ is small.
Close to the center frequency, strong reflections occur, and for transducers with a large number of cells, we have the approximate relation $P_{11} = P_{22} \approx i$ (see Ref.~\onlinecite{2007Morg}, Eq.~(9.8)). The delay lines measured in the main paper have between 125 and 135 cells, and fits to the data give $P_{11} \approx P_{22} \approx 0.8 \, i$, reasonably close to the approximation. Introducing this approximation in the transmission ratio yields
\begin{align}\label{eq:5}
    \frac{S_{21}^\mathrm{\rightarrow \leftarrow}}{S_{21}^\mathrm{\leftarrow \rightarrow}} & \approx \frac{\left(P_{13}\right)^2 \left( \left(P_{23}\right)^4 - \left( i \left(P_{23}\right)^2 -  P_{33} -  Y_{0}\right)^{2}\right)}{\left(P_{23}\right)^2 \left( \left(P_{13}\right)^4 - \left( i \left(P_{13}\right)^2 -  P_{33} -  Y_{0}\right)^{2}\right)}.
\end{align}
Taking the norm of Eq.~(\ref{eq:5}),
\begin{align}
    \left |\frac{S_{21}^\mathrm{\rightarrow \leftarrow}}{S_{21}^\mathrm{\leftarrow \rightarrow}}\right| & = \left|\frac{P_{13} }{P_{23}}\right|^2 \left |\frac{\left(P_{23}\right)^4 - \left( i \left(P_{23}\right)^2 - P_{33} - Y_{0}\right)^{2}}{ \left(P_{13}\right)^4 - \left( i \left(P_{13}\right)^2 - P_{33} - Y_{0}\right)^{2}}\right |.
\end{align}
We rearrange terms to give
\begin{align}
    \left |\frac{S_{21}^\mathrm{\rightarrow \leftarrow}}{S_{21}^\mathrm{\leftarrow \rightarrow}}\right| & = \left|\frac{P_{13} }{P_{23}}\right|^2 \left |\frac{ i \left(P_{23}\right)^2 - P_{33} - Y_{0}}{i \left(P_{13}\right)^2 - P_{33} - Y_{0}}\right |^{2} \left |\frac{\left(\frac{\left(P_{23}\right)^2}{i \left(P_{23}\right)^2 - P_{33} - Y_{0}}\right)^2 - 1}{ \left(\frac{\left(P_{13}\right)^2}{i \left(P_{13}\right)^2 - P_{33} - Y_{0}}\right)^2 - 1}\right |.
\end{align}
Noting that $|i (P_{13|23})^2 -P_{33} - Y_{0}| \approx |-P_{33} - Y_{0}|$, we obtain
\begin{align}
    \left |\frac{S_{21}^\mathrm{\rightarrow \leftarrow}}{S_{21}^\mathrm{\leftarrow \rightarrow}}\right | & \approx \left |\frac{P_{13} }{P_{23}}\right |^2.
\end{align}

Using the definition of the directivity from Ref.~\onlinecite{2007Morg}, we conclude
\begin{equation}\label{eq:final-d}
    D = \left|\frac{P_{13} }{P_{23}}\right| \approx \sqrt{\left|\frac{S_{21}^\mathrm{\rightarrow \leftarrow}}{S_{21}^\mathrm{\leftarrow\rightarrow} }\right|}.
\end{equation}

\begin{figure}[htbp]
    \centering
    \includegraphics[width=0.6\linewidth]{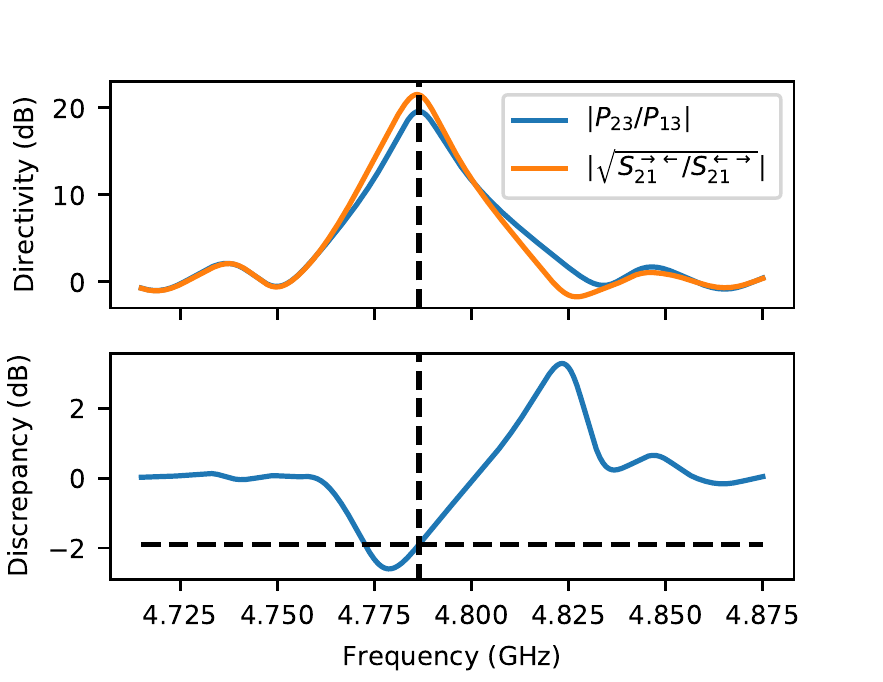}
    \caption{Comparison of the two expression of Eq.~\ref{eq:final-d}, top panel, and their difference, bottom panel.}
    \label{fig:difference}
\end{figure}

As a check, we can use our circuit model to generate directivities based on the two expressions. In Fig.~\ref{fig:difference} we compare the two expressions in the top panel, and show the difference in the bottom panel. We note that when we use the circuit parameters from fits to the experimental data, the difference between the two expressions at the center frequency is $\approx$ \SI{2}{\decibel}, which translates to $\approx$ \SI{24}{\percent} relative error.

\bibliographystyle{apsrev4-1}
\bibliography{bib_supp}